\documentclass{article}
\usepackage{Odyssey2016,epsfig,amssymb,amsmath}
\usepackage{graphicx,mathtools,mathalfa,multirow,balance,url}
\usepackage{epstopdf}
\ninept

\DeclareMathOperator*{\argmax}{arg\,max}

\DeclareMathOperator{\tr}{tr}

\title{The IBM 2016 Speaker Recognition System}

\name{Seyed Omid Sadjadi$^1$, Sriram Ganapathy$^{2\star}$\thanks{$^\star$This work was done while Sriram Ganapathy was a Research Staff Member at IBM.}, Jason W. Pelecanos$^1$}

\address{$^1$IBM Research, Yorktown Heights, NY, USA \\
	$^2$Dept. of Electrical Eng., Indian Institute of Science, Bangalore, India \\
{\small \tt sadjadi@us.ibm.com} }
\begin{document}
\maketitle
\begin{abstract}
In this paper we describe the recent advancements made in the IBM i-vector speaker recognition system for conversational speech. In particular, we identify key techniques that contribute to significant improvements in performance of our system, and quantify their contributions. The techniques include: 1) a nearest-neighbor discriminant analysis (NDA) approach that is formulated to alleviate some of the limitations associated with the conventional linear discriminant analysis (LDA) that assumes Gaussian class-conditional distributions, 2) the application of speaker- and channel-adapted features, which are derived from an automatic speech recognition (ASR) system, for speaker recognition, and 3) the use of a deep neural network (DNN) acoustic model with a large number of output units ($\sim~\!\!10$k senones) to compute the frame-level soft alignments required in the i-vector estimation process. We evaluate these techniques on the NIST 2010 speaker recognition evaluation (SRE) extended core conditions involving telephone and microphone trials. Experimental results indicate that: 1) the NDA is more effective (up to 35\% relative improvement in terms of EER) than the traditional parametric LDA for speaker recognition, 2) when compared to raw acoustic features (e.g., MFCCs), the ASR speaker-adapted features provide gains in speaker recognition performance, and 3) increasing the number of output units in the DNN acoustic model (i.e., increasing the senone set size from 2k to 10k) provides consistent improvements in performance (for example from 37\% to 57\% relative EER gains over our baseline GMM i-vector system). To our knowledge, results reported in this paper represent the best performances published to date on the NIST SRE 2010 extended core tasks.             
\end{abstract}

\section{Introduction}
There have been significant advancements in the speaker recognition field over the past few years. The research trend in this field has gradually evolved from joint factor analysis (JFA) based methods, which attempt to model the speaker and channel subspaces separately \cite{Kenny2007}, towards the i-vector approach that models both speaker and channel variabilities in a single low-dimensional (e.g., a few hundred) space termed the total variability subspace \cite{Dehak2011}. State-of-the-art i-vector based speaker recognition systems employ universal background models (UBM) to generate frame-level soft alignments required in the i-vector estimation process. The i-vectors are typically post-processed through a linear discriminant analysis (LDA) \cite{Fukunaga1990} stage to generate dimensionality reduced and channel-compensated features which can then be efficiently modeled and scored with various backends such as a probabilistic LDA (PLDA) \cite{Prince2007, Garcia2011}. 

Until recently, Gaussian mixture models (GMM) trained in an unsupervised fashion (i.e., with no phonetic labels) were commonly used to represent the UBM in speaker recognition. However, inspired by the success of deep neural network (DNN) acoustic models in the automatic speech recognition (ASR) field, \cite{Lei2014a} proposed the use of DNN senone (context-dependent triphones) posteriors for computing the soft alignments, which resulted in remarkable reductions in speaker recognition error rates. The performance improvements reported in \cite{Lei2014a} are consistent with the observations made in our earlier effort \cite{Omar2010} where a supervised GMM-HMM acoustic model (derived from an ASR system) was utilized to estimate the hyperparameters of a phonetically inspired UBM (PI-UBM) for speaker recognition. More recently, a supervised GMM-UBM (with full covariance matrices) based on DNN posteriors was also successfully evaluated for telephony speaker recognition \cite{Snyder2015}. These approaches are motivated by the fact that many of the speaker-dependent characteristics, which are conditioned on some phonetic units/classes, may be more effectively modeled using a UBM trained with explicit phonetic information. 


In this paper, we report on the latest advancements made in the IBM i-vector speaker recognition system for conversational speech. Particularly, we first describe key components that contribute to significant improvements in performance of our system. These components include: 1) a nearest-neighbor based discriminant analysis (NDA) approach \cite{Sadjadi2014} for channel compensation in i-vector space, which, unlike the commonly used Fisher LDA, is non-parametric and typically of full rank, 
2) speaker- and channel-adapted features derived from feature-space maximum likelihood linear regression (fMLLR) transforms \cite{Digalakis1995, Gales1998}, which are used both to train/evaluate the DNN and to compute the sufficient Baum-Welch statistics for i-vector extraction, 
and 3) a DNN acoustic model with a large number of output units ($\sim~\!\!\!10$k~senones) to compute the soft alignments (i.e., the posteriors). To quantify the contribution of these components, we evaluate our system in the context of speaker verification experiments using speech material from the NIST 2010 speaker recognition evaluation (SRE) which includes 5 extended core conditions involving telephone and microphone trials.

\begin{figure*}[t]
	\centering
	\includegraphics[scale=.55, clip, trim=0mm 5mm 7mm 0mm] {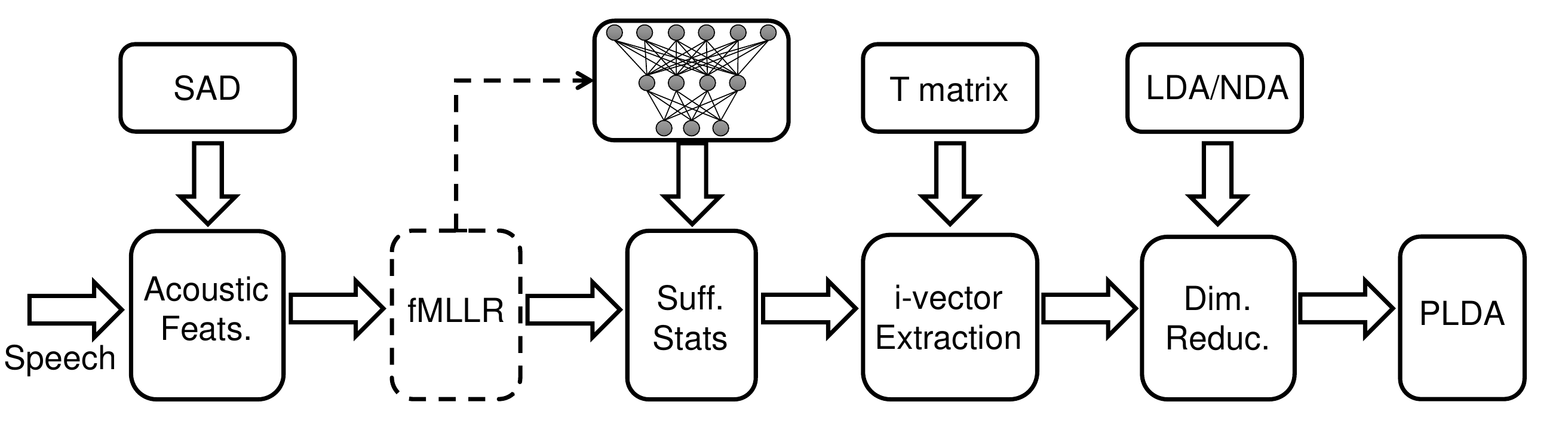}
	\vspace{-0mm}
	\caption{\it Block diagram of the IBM speaker recognition system with fMLLR speaker- and channel-adapted features, DNN posterior based i-vectors, and NDA dimensionality reduction.}
	\label{fig:blk}
	\vspace{-0mm}
\end{figure*}

\section{System Overview}

In the following subsections, we briefly describe the major components of our speaker recognition system. Specifically, we first provide an overview of GMM- versus DNN-based i-vector extraction, which is followed by algorithmic descriptions for the LDA and the NDA for channel compensation in the i-vector space. A schematic block diagram of the system is depicted in Fig.~\ref{fig:blk}.   

\subsection{I-vector extraction}

The i-vector representation is based on the total variability modeling concept which assumes that speaker- and channel-dependent variabilities reside in the same low-dimensional subspace \cite{Dehak2011}. The key idea here is that variability within and across sessions can be described via a small set of parameters (a.k.a factors) in a low-dimensional subspace spanned by the columns of a low-rank rectangular matrix, $\mathbf{T}$, entitled the \textit{total variability matrix}. Mathematically, the adapted mean supervector, $\mathbf{M}$(s), for a given set of observations, $s$, can be modeled as,
\begin{equation}
\mathbf{M}(s) = \mathbf{m} + \mathbf{T}\,\mathbf{x}(s) + \boldsymbol{\epsilon},
\end{equation}  
where $\mathbf{m}$ is the prior mean supervector, $\mathbf{x}(s)\sim \mathcal{N} (\mathbf{0},\mathbf{I})$ is a multivariate random variable termed an identity vector ``i-vector", and $\boldsymbol{\epsilon}\sim \mathcal{N} (\mathbf{0},\boldsymbol{\Sigma})$ is a residual noise term to account for the variability not captured via $\mathbf{T}$ ($\boldsymbol\Sigma$ is typically copied from the UBM). In other words, for the given observation set $s$, the i-vector represents the coordinates in the total variability subspace.

In order to learn the bases for the total variability subspace, one needs to compute the Baum-Welch statistics which are defined as,
\begin{eqnarray}
\label{eqn:zeroth}
N_g(s) &=& \sum_{t}\gamma_{tg}(s),\\
\label{eqn:first}
\mathbf{F}_g(s) &=& \sum_{t}\gamma_{tg}(s)\, \mathbf{O}_t(s),
\end{eqnarray}
where $N_g(s)$ and $\mathbf{F}_g(s)$ denote the zeroth- and first-order statistics for speech session $s$, respectively, with $\gamma_{tg}(s)$ being the posterior probability of the mixture component $g$ given the observation vector $\mathbf{O}_t(s)$ at time frame $t$. 

The observation vector $\mathbf{O}_t(s)$ can be either the conventional raw acoustic features such as mel-frequency cepstral coefficients (MFCC) or their speaker- and channel-adapted versions which is computed through a per recording fMLLR transform \cite{Gales1998, Digalakis1995} typically obtained with a GMM-HMM system. Note from Fig.~\ref{fig:blk} that the same fMLLR transformed features can be used to train/evaluate the DNN as well as compute the sufficient Baum-Welch statistics for i-vector extraction.


Traditionally, the frame-level soft alignments, $\gamma_{tg}(s)$, in (\ref{eqn:zeroth}) and (\ref{eqn:first}) are computed with a GMM acoustic model trained in an unsupervised fashion (i.e., with no phonetic labels). However, in \cite{Omar2010}, a supervised GMM-HMM acoustic model (derived from a speech recognition system) was utilized to estimate the GMM-UBM hyperparameters for speaker recognition, assuming that class-conditional distributions for the various phonetic classes are Gaussian. More recently, inspired by the success of DNN acoustic models in automatic speech recognition (ASR) field, \cite{Lei2014a} proposed the use of DNN senone (context-dependent triphones) posteriors for computing the soft alignments, $\gamma_{tg}(s)$, which resulted in remarkable reductions in speaker recognition error rates. Motivated by these results, in this paper, we explore the DNN senone posterior based i-vectors for speaker recognition, and compare their effectiveness against GMM i-vectors on this task.  Furthermore, we also investigate the impact of the senone set size on speaker recognition performance. It is worth noting that increasing the number of components in the unsupervised GMM acoustic model (with diagonal covariance matrices) for speaker recognition did not seem to result in much performance gains, if at all, in the recent studies \cite{Lei2014a, Snyder2015}. 

\subsection{Linear discriminant analysis (LDA)}
As noted before, i-vectors model speaker- and channel-dependent information within the same total variability subspace. Therefore, in order to select the most relevant feature subset for the speaker recognition task, LDA can be applied to i-vectors to annihilate the directions not informative for speaker recognition. In addition, reducing the dimensionality of i-vectors via LDA can improve the computational efficiency of the subsequent backend components in the system. 

LDA computes an optimum linear projection $\mathbf{A}\!\!: \mathbb{R}^d \mapsto \mathbb{R}^n$,
by maximizing the ratio of the inter-class scatter to intra-class variance, where $\bf{A}$ is a rectangular matrix with $n$ linearly independent columns. Here, the within- and between-class scatter matrices are used to formulate a class separability criterion which converts the matrices into a single statistic. This statistic takes on larger values when the between-class scatter is larger and the within-class variance is smaller. Several such class separability criteria are described in \cite{Fukunaga1990}, of which the following is the most widely used,
\begin{equation} \label{eqn:ray}
\hat{\mathbf{A}} = \argmax_{\mathbf{A}^T\mathbf{S}_{w}\mathbf{A} = \mathbf{I}} \left[\tr\left(\mathbf{A}^T\mathbf{S}_{b}\mathbf{A}\right)\right],
\end{equation}    
where $\mathbf{S}_b$ and $\mathbf{S}_w$ denote the between- and within- class scatter matrices, respectively. The optimization problem in (\ref{eqn:ray}) has an analytical solution that is a matrix whose columns are the $n$ eigenvectors corresponding to the largest eigenvalues of $\mathbf{S}_w^{-1}\mathbf{S}_b$.

There are three disadvantages associated with the parametric nature of the scatter matrices$\mathbf{S}_b$ and $\mathbf{S}_w$. First, the underlying distribution of classes is assumed to be Gaussian with a common covariance matrix for all classes. Therefore, one cannot expect the parametric LDA to generalize well to non-Gaussian and multi-modal (as opposed to unimodal) distributions. It is well known in the speaker recognition community that the actual distribution of i-vectors may not necessarily be Gaussian \cite{Kenny2010}. This is in particular more problematic when speech recordings are collected in the presence of noise and channel distortions \cite{Sadjadi2014, Sadjadi2015}. In addition, for the NIST SRE type of scenarios, speech recordings come from various sources and collects (sometimes out-of-domain), therefore unimodality of the distributions cannot be guaranteed.  Second, notice that the rank of $\mathbf{S}_b$ is $C-1$, which means the parametric LDA can provide at most $C-1$ discriminant features. However, this may not be sufficient in applications such as language recognition where the number of language classes is much smaller than the dimensionality of the i-vectors \cite{Sadjadi2015}. Nevertheless, this may not pose a challenge for speaker recognition tasks in which the number of training speakers exceeds the dimensionality of the total variability subspace. Finally, because only the class centroids are taken into account for computing $\mathbf{S}_b$, the parametric LDA cannot effectively capture the boundary structure between adjacent classes which is essential for classification \cite{Fukunaga1990}.

To overcome the above noted limitations of LDA, an NDA technique was proposed in \cite{Fukunaga1983}, that measures both the within- and between-class scatters on a local basis using a nearest neighbor rule. We have previously evaluated the NDA for both speaker and language recognition tasks on high-frequency (HF) radio channel degraded data \cite{Sadjadi2014, Sadjadi2015}, where it compared favorably to the LDA. We provide a brief description of NDA in the next section.


\subsection{Nearest-neighbor discriminant analysis (NDA)}

\begin{figure}[t]
	\centering
	\includegraphics[scale=.6, clip, trim=80mm 60mm 3mm 3mm] {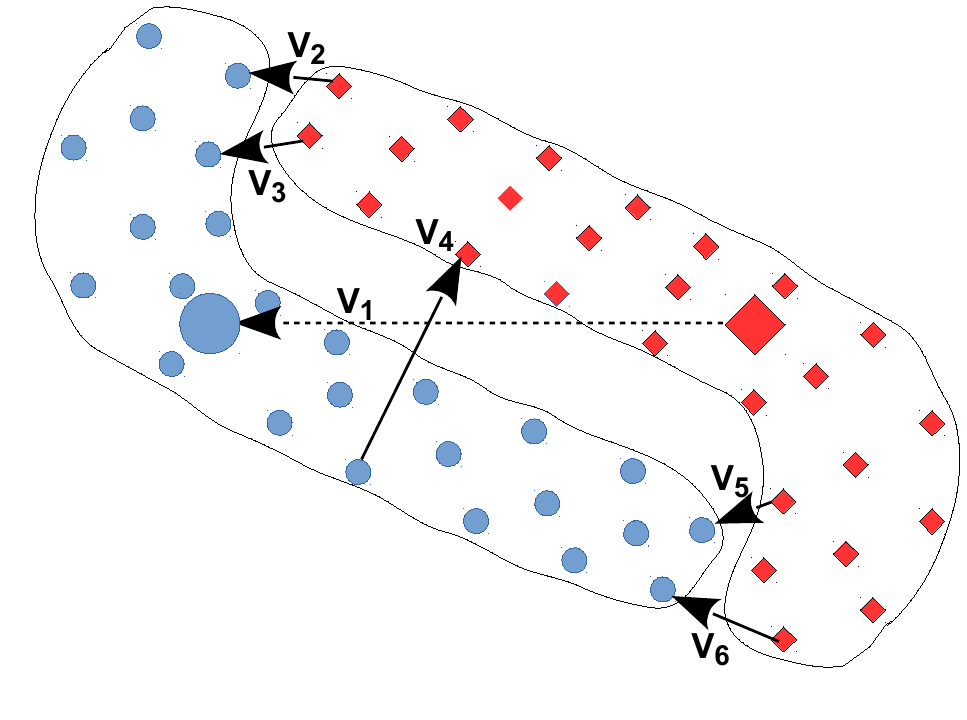}
	\caption{{\it Symbolic example illustrating the parametric versus nonparametric scatter between two classes. $\mathrm{v}_1$ represents the global gradient of class centroids. The vectors $\{\mathrm{v}_2, \cdots, \mathrm{v}_6\}$ represent the local gradients.}}
	\label{fig:fig1}
\end{figure}

In order to alleviate some of the limitations identified for LDA, a nonparametric discriminant analysis techniques was proposed in \cite{Fukunaga1983}. In NDA, the expected values that represent the global information about each class are replaced with local sample averages computed based on the $k$-NN of individual samples. More specifically, in the NDA approach, the between-class scatter matrix is defined as,
\begin{equation} \label{eqn:betweenk}
\tilde{\mathbf{S}}_b = \sum_{i=1}^{C} \sum_{\substack{j=1 \\ j\neq i}}^{C} \sum_{l=1}^{N_i} w^{ij}_l {\left(\mathbf{x}^i_l-\mathcal{M}^{ij}_l\right)\left(\mathbf{x}^i_l-\mathcal{M}^{ij}_l\right)^T},
\end{equation}     
where $\mathbf{x}^i_l$ denotes the $l^\textrm{th}$ sample from class $i$, and $\mathcal{M}^{ij}_l$ is the local mean of $k$-NN samples for $\mathbf{x}^i_l$ from class $j$ which is computed as,
\begin{equation}
\mathcal{M}^{ij}_l = \frac{1}{K}\sum_{k=1}^{K} NN_k(\mathbf{x}^i_l,j),
\end{equation}
where $NN_k(\mathbf{x}^i_l,j)$ is the $k^\textrm{th}$ nearest neighbor of $\mathbf{x}^i_l$ in class $j$. The weighting function $w^{ij}_l$ in (\ref{eqn:betweenk}) is defined as,
\begin{equation}\label{eqn:wght}
w^{ij}_l = \frac{\min\left\{d^\alpha\!\!\left(\mathbf{x}^i_l, NN_K(\mathbf{x}^i_l,i)\right),d^\alpha\!\!\left(\mathbf{x}^i_l, NN_K(\mathbf{x}^i_l,j)\right)\right\}}{d^\alpha\!\!\left(\mathbf{x}^i_l, NN_K(\mathbf{x}^i_l,i)\right)+d^\alpha\!\!\left(\mathbf{x}^i_l, NN_K(\mathbf{x}^i_l,j)\right)},
\end{equation}
where $\alpha\!\in\! \mathbb{R}$ is a constant between zero and infinity, and $d(.)$ denotes the distance (e.g., cosine or Euclidean). The weighting function is introduced in (\ref{eqn:betweenk}) to deemphasize the local gradients that are large in magnitude to mitigate their influence on the scatter matrix. The weight parameters approach $0.5$ for samples near the classification boundary (e.g., see $\{\mathrm{v}_2, \mathrm{v}_3, \mathrm{v}_5, \mathrm{v}_6\}$ shown in Figure~\ref{fig:fig1}), while dropping off to $0$ for samples that are far from the boundary (e.g., see $\mathrm{v}_4$ in Figure~\ref{fig:fig1}). The control parameter $\alpha$ determines how rapidly such decay in the weights occurs. 

The nonparametric within-class scatter matrix, $\tilde{\mathbf{S}}_w$, is computed in a similar fashion as in (\ref{eqn:betweenk}), except the weighting function is set to $1$ and the local gradients are computed within each class. The NDA transform is then formed by calculating the eigenvectors of $\tilde{\mathbf{S}}_w^{-1}\tilde{\mathbf{S}}_b$.

Three important observations can be made from a careful examination of the nonparametric between-class scatter matrix in (\ref{eqn:betweenk}). First, notice that as the number of nearest neighbors, $K$, approaches $N_j$, the total number of samples in class $j$, the local mean vector, $\mathcal{M}^{ij}_l$, approaches the global mean of class $j$ (i.e., $\boldsymbol{\mu}_j$). In this scenario, if we set the weight parameters to $1$, the NDA transform essentially becomes the LDA projection, which means the LDA is a special case of the more general NDA.

Second, because all the samples are taken into account for the calculation of the nonparametric between-class scatter matrix (as opposed to only the class centroids), $\tilde{\mathbf{S}}_b$ is generally of full rank. This means that unlike the LDA that provides at most $C-1$ discriminant features, the NDA generally results in $d$-dimensional vectors (assuming a $d$-dimensional input space) for the classification. As we discussed before, this is of great importance for applications such as language recognition where the number of classes is much smaller than the dimensionality of the total subspace (or the input space in general). 

Finally, compared to LDA, NDA is more effective in preserving the complex structure (i.e., local and boundary structure) within and across different classes. As seen from the example shown in Figure~\ref{fig:fig1} (where $k$ is set to $1$ for simplicity), LDA only uses the global gradient obtained with the centroids of the two classes (i.e., $\mathrm{v}_1$) to measure the between-class scatter. On the other hand, NDA uses the local gradients (i.e., $\{\mathrm{v}_2, \cdots, \mathrm{v}_6\}$) that are emphasized along the boundary through the weighting function, $w^{ij}_l$. Hence, the boundary information becomes embedded into the resulting transformation.

\section{Experiments}
\label{sec:expts}

This section provides a description of our experimental setup including speech data, the ASR system configuration, and the speaker recognition (SR) system configuration used in our evaluations. 

\begin{table*}[th]
	\renewcommand{\tabcolsep}{1.5 mm} 
	\renewcommand{\arraystretch}{1.5} 
\caption{\label{tab:tab1} {\it Description of the 5 core enrollment/test conditions in the NIST 2010 SRE.}}
\vspace{2mm}
\centerline{
\begin{tabular}{|l|l|l|c|c|c|}
\hline
Condition & Enroll & Test & Mismatch & \#Target Trials & \#Impostor Trials \\
\hline  \hline
C1 & Interview microphone & Interview microphone (same type) & No & 4,034 & 795,995 \\
C2 & Interview microphone & Interview microphone (different type) & Yes & 15,084 & 2,789,534 \\
C3 & Interview microphone & Telephony & Yes & 3,989 & 637,850 \\
C4 & Interview microphone & Room microphone & Yes & 3,637 & 756,775 \\
C5 & Telephony & Telephony (different type) & Yes & 7,169 & 408,950\\
\hline
\end{tabular}}
\end{table*}

\begin{figure}[!b]
	\centering
	\includegraphics[scale=0.97, angle=-90, clip, trim=7mm 5mm 5mm 5mm] {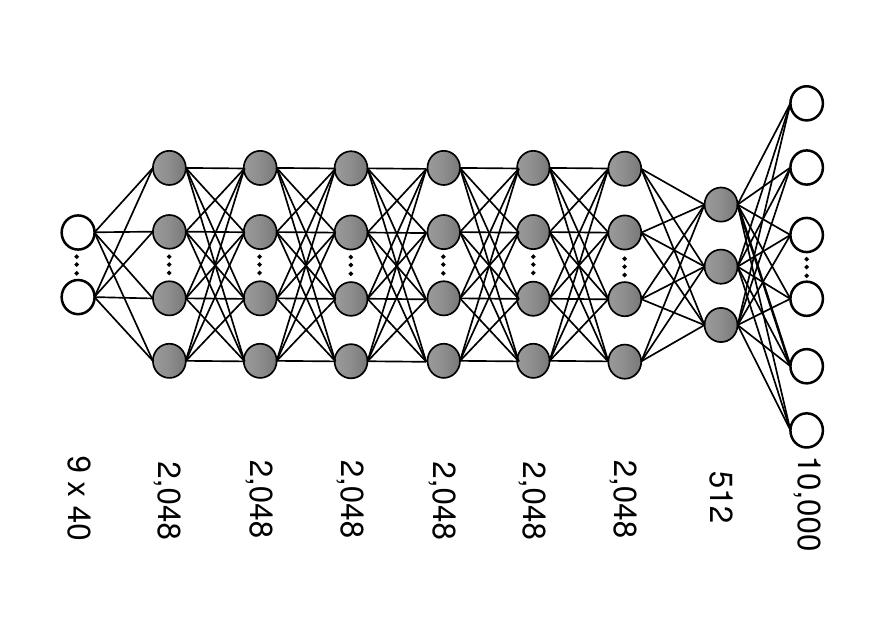}
	\vspace{-0mm}
	\caption{\it Architecture of the DNN acoustic model with 7 hidden layers used in our speaker recognition experiments.}
	\label{fig:dnn}
	\vspace{-0mm}
\end{figure}

\subsection{Data}
We conduct the core of our speaker recognition experiments using conversational telephone and microphone (phone call and interview) speech material extracted from datasets released through the linguistic data consortium (LDC) for the NIST 2004-2010 SRE \cite{Cieri2007, Brandschain2010}, as well as Switchboard Cellular (SWBCELL) Parts I and II and Switchboard2 (SWB2) Phase II and Phase III corpora. These datasets contain speech spoken in U.S. English (the non English portion was filtered out) from a large number of male and female speakers with multiple sessions per speaker. The NIST SRE 2010 data is held out for evaluations, while the remaining data are used to train the system hyper-parameters (i.e., the i-vector extractor, LDA/NDA, and PLDA). There are a total of 5 extended core tasks in the NIST SRE 2010 that involve telephone and microphone trials from both male and female speakers \cite{SRE2010}. A more detailed description of the 5 tasks is presented in Table~\ref{tab:tab1}.  

\subsection{DNN system configuration}
The architecture of the DNN acoustic model used to generate the soft alignments for i-vector extraction is shown in Fig.~\ref{fig:dnn}. The model, which has 7 fully connected hidden layers with 2048 units per layer except for the bottleneck layer that has 512 units, is discriminatively trained using the standard error back-propagation and cross-entropy objective function to estimate posterior probabilities of 10,000 senones (HMM triphone states). The training is accomplished using the IBM Attila toolkit \cite{Soltau2010} on 600 hours of conversational telephone speech (CTS) data from the Fisher corpus \cite{Cieri2004} with a 9-frame context of 40-dimensional speaker-adapted feature vectors obtained through per recording fMLLR transforms \cite{Digalakis1995, Gales1998}. The fMLLR transforms are generated for each recording with decoding alignments obtained from a GMM-HMM acoustic model. The GMM models are trained with 40-dimensional features which are derived from 13-dimensional MFCCs as follow; the base cepstral features from 9 consecutive frames are first spliced after cepstral mean-variance and vocal tract length normalizations (VTLN). An LDA transform is then applied to reduce the final feature vector dimensionality to 40. The range of the LDA transformation is diagonalized by means of a global semi-tied covariance transform (see \cite{Ganapathy2015, Saon2015} for more details). In addition to running experiments with all the 10k senones, we also explore smaller senone set sizes of 2k and 4k which are obtained by merging the 10k HMM states using a phonetic decision tree with maximum-likelihood (ML) criterion \cite{Young1994}.

\subsection{SR system configuration}
For speech parameterization (other than the fMLLR based features), we extract 13-dimensional MFCCs (including $c_0$) from 25~ms frames every 10~ms using a 24-channel mel filterbank spanning the frequency range 200-3500 Hz. The first and second temporal cepstral derivatives are also computed over a 5-frame window and appended to the static features to capture the dynamic pattern of speech over time. This results in 39-dimensional feature vectors. For non-speech frame dropping, we employ an unsupervised speech activity detector (SAD) based on voicing energy features \cite{Sadjadi2013}. After dropping the non-speech frames, global (recording level) cepstral mean and variance normalization (CMVN) is applied to suppress the short term linear channel effects.

In this paper, a 500-dimensional total variability subspace is learned and used to extract i-vectors from the recordings.  To learn the i-vector extractor, out of a total of 60,178 recordings available from 1884 male and 2601 female speakers, we select 48,325 recordings from the NIST SRE 2004-2008, SWBCELL, and SWB2 corpora. The zeroth and first order Baum-Welch statistics are computed for each recording using soft alignments obtained from either a gender-independent 2048-component GMM-UBM with diagonal covariance matrices, or the DNN acoustic model with 2k, 4k, and 10k senones. The GMM-UBM is trained using 21,207 recordings selected from the NIST SRE 2004-2006, SWBCELL, and SWB2 corpora.

After extracting 500-dimensional i-vectors, we either use LDA or NDA for inter-session variability compensation by reducing the dimensionality to 250. In order to train the NDA, we employ a one-versus-rest strategy to compute the inter-speaker scatter matrix in (\ref{eqn:betweenk}). This provides flexibility on the number of nearest neighbors used for computing the local means. A cosine similarity metric (as opposed to Euclidean) is used to find the $k$-nearest neighbors for each sample, and the exponent $\alpha$ in (\ref{eqn:wght}) is set to $1$. The dimensionality reduced i-vectors are then centered (the mean is removed), whitened, and unit-length normalized. For scoring, a Gaussian PLDA model with a full covariance residual noise term \cite{Prince2007, Garcia2011} is learned using the i-vectors extracted from all 60,178 speech segments (1884 male and 2601 female speakers) as noted previously. The Eigenvoice subspace in the PLDA model is assumed full-rank. 

\begin{figure}[!b]
	\centering
	\includegraphics[scale=0.48, clip, trim=1mm 2mm 1mm 1mm] {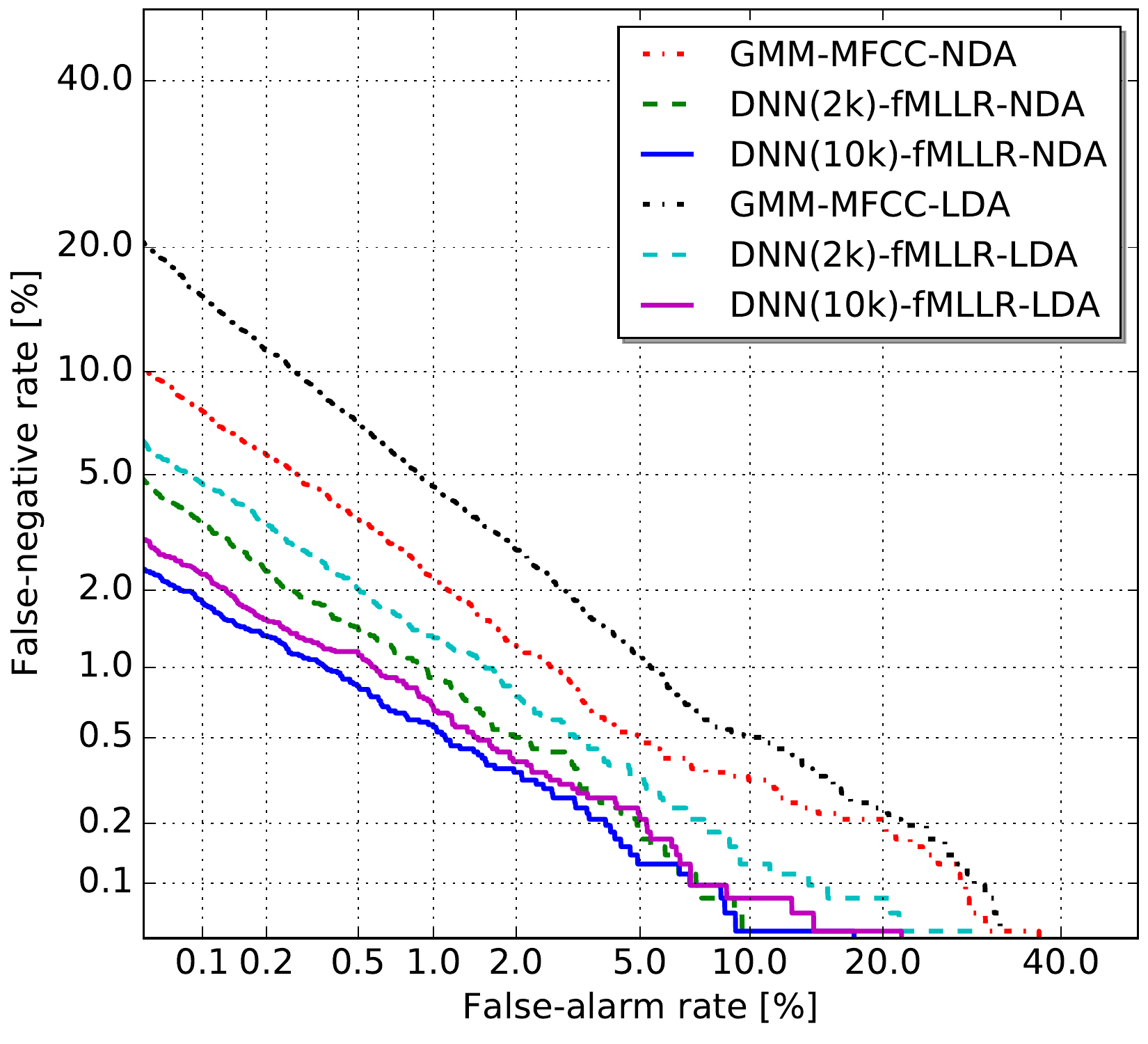}
	\vspace{-3mm}
	\caption{\it DET plot comparison of IBM speaker recognition systems with various configurations on extended core condition 5 in the NIST SRE 2010.}
	\label{fig:det}
	\vspace{-0mm}
\end{figure}

\begin{table}[t]
	\renewcommand{\tabcolsep}{1.5 mm} 
	\renewcommand{\arraystretch}{1.5} 
	\caption{\label{tab:tab2} {\it Performance comparison of IBM speaker recognition systems with various configurations on extended core condition 5 in the NIST SRE 2010. A DNN with 10k senones is used.}}
	\vspace{2mm}
	\centerline{
		\begin{tabular}{|l|c|c|c|}
			\hline
			System & EER [\%] & minDCF08 & minDCF10 \\
			\hline  \hline
			GMM-MFCC-LDA & 2.40  & 0.120 & 0.439 \\
			GMM-MFCC-NDA & 1.55 & 0.076 & 0.286 \\
			DNN-MFCC-LDA & 1.02 & 0.045 & 0.168 \\
			DNN-MFCC-NDA & 0.76 & 0.036 & 0.147 \\
			DNN-fMLLR-LDA & 0.82 & 0.032 & 0.120 \\
			DNN-fMLLR-NDA & \textbf{0.67} & \textbf{0.028} & \textbf{0.092} \\
			\hline
		\end{tabular}}
\end{table}

\begin{table}[t]
	\renewcommand{\tabcolsep}{1.2 mm} 
	\renewcommand{\arraystretch}{1.5} 
	\caption{\label{tab:tab3} {\it Performance comparison of IBM speaker recognition systems with fMLLR features for 2k, 4k, and 10k DNN senones on extended core condition 5 in the NIST SRE 2010.}}
	\vspace{2mm}
	\centerline{
		\begin{tabular}{|l|c|c|c|c|}
			\hline
			System & \#Senones & EER [\%] & minDCF08 & minDCF10 \\
			\hline  \hline
			DNN-LDA & 2k & 1.19  & 0.054 & 0.212 \\
			DNN-NDA & 2k & 0.95 & 0.043 & 0.166 \\
			DNN-LDA & 4k & 0.98 & 0.041 & 0.169 \\
			DNN-NDA & 4k & 0.86 & 0.033 & 0.116 \\
			DNN-LDA & 10k & 0.82 & 0.032 & 0.120 \\
			DNN-NDA & \textbf{10k} & \textbf{0.67} & \textbf{0.028} & \textbf{0.092} \\
			\hline
		\end{tabular}}
\end{table}

\begin{table*}[!t]
	\renewcommand{\tabcolsep}{2.3 mm} 
	\renewcommand{\arraystretch}{1.8} 
	\caption{\it Performance comparison of IBM speaker recognition systems with various configurations on extended microphone and telephone conditions (C1--C4) in the NIST SRE 2010. A DNN model with 10k senones is used.}
	\vspace{-1mm}
	\begin{center}
		\begin{tabular}{ lcccc|cccc|cccc }
			\hline\hline
			\multirow{2}{*}{System} & \multicolumn{4}{c|}{EER [\%]} & \multicolumn{4}{c|}{minDCF08} & \multicolumn{4}{c}{minDCF10}\\ 
			\cline{2-13}
			{ } & {C1} & {C2} & {C3} & {C4} & {C1} & {C2} & {C3} & {C4} & {C1} & {C2} & {C3} & {C4}\\
			\hline \hline
			{GMM-MFCC-NDA} & 1.39 & 1.89 & 1.80 & 1.46 & 0.053 & 0.084 & 0.081 & 0.061 & 0.215 & 0.313 & 0.315 & 0.251 \\
			{DNN-MFCC-NDA} & \textbf{0.84} & \textbf{1.41} & \textbf{0.83} & \textbf{0.63} & \textbf{0.027} & \textbf{0.046} & 0.036 & \textbf{0.022} & \textbf{0.104} & \textbf{0.157} & 0.127 & 0.103 \\
			{DNN-fMLLR-NDA} & 1.02 & 1.44 & 0.90 & 0.77 & 0.033 & 0.049 & \textbf{0.034} & 0.025 & 0.112 & 0.158 & \textbf{0.119} & \textbf{0.096} \\
			\hline\hline
		\end{tabular}
	\end{center}
	\label{tab:tab4}
\end{table*}

\section{Results and Discussion}
\label{sec:result}

In this section, we summarize our results obtained with the experimental setup presented in Section~\ref{sec:expts}. In the first experiment, we evaluated the effectiveness of the NDA versus the LDA for inter-session variability compensation and dimensionality reduction in the i-vector space. The outcome of this experiment is presented in Table~\ref{tab:tab2} for the NIST SRE 2010 extended ``tel-tel'' trials (condition 5), in terms of the equal error rate (EER), minimum detection cost function with the NIST SRE 2008 \cite{SRE2008} and 2010 \cite{SRE2010} definitions (minDCF08 and minDCF10). It can be seen from the table that the systems with the NDA consistently provide better speaker recognition performance across all three metrics. For the GMM based system, a relative improvement of 35\% in EER is achieved with the NDA over the LDA, while for the DNN based systems with MFCCs and fMLLR features relative improvements of 26\% and 18\% are obtained, respectively. As we discussed before, this is due to the nonparametric representations for the scatter matrices in NDA that makes no assumptions regarding the underlying class-conditional distributions. In addition, NDA is more effective in capturing the local structure (as opposed to global bulk structure) and boundary information within and across different speakers. Another important observation that can be made from Table~\ref{tab:tab2} is that, irrespective of the dimensionality reduction algorithm used, the systems with fMLLR features outperform the MFCC based systems. This is attributed to the ability of the fMLLR transforms in reducing the speaker and channel variabilities in the acoustic feature space.  

In the next set of experiments, we investigated the impact of the number of senones on speaker recognition performance. Table~\ref{tab:tab3} shows speaker recognition results on the NIST SRE 2010 ``tel-tel'' condition which are obtained with i-vectors computed using 2k, 4k, and 10k DNN senones and fMLLR features. Two important observations can be made from this table. First, the larger the number of senones, the better the performance. This is due to the discriminative nature of the DNN where increasing the granularity in the output layer improves the model ability in distinguishing among the various phonetic events. It is worth noting that increasing the number of components in the unsupervised GMM acoustic model (with diagonal covariance matrices) for speaker recognition did not result in much performance improvements in the recent studies \cite{Lei2014a, Snyder2015}. Second, irrespective of the number of senones used to calculate the sufficient statistics, the NDA based systems consistently perform better than the LDA based systems. We note that, in our experiments, increasing the number of senones beyond 10k did not yield much gains in performance.

Fig.~\ref{fig:det} shows the detection error trade-off (DET) curves for the NDA and LDA based systems on the extended core condition 5 in the NIST SRE 2010. Consistent with our previous observations, it is seen that the NDA based systems achieve the best performance across a wide range of operating points on the DET curves. The performance gap between the NDA and LDA based systems is, however, reduced when DNN senone posteriors are used to compute the i-vectors, and increasing the senone set size from 2k to 10k further narrows this gap.      

For completeness, we also evaluated the performance of our speaker recognition system on extended microphone and telephone conditions (C1--C4) in the NIST SRE 2010. The results are provided in Table~\ref{tab:tab4} for both the GMM and DNN based systems. It is clear that the DNN based systems, with either MFCCs or fMLLR features, perform significantly better than the GMM based system. Additionally, the DNN based system trained with raw MFCCs tend to perform better than the fMLLR based system, at least in terms of EER. We speculate that this is because the fMLLR transforms, which are obtained using GMM-HMMs trained only on telephony data, are unable to effectively reduce the variability due to channel mismatch on microphone recordings.  

\section{Conclusions}
In this paper, we presented the recent improvements made in our state-of-the-art i-vector speaker recognition system. We investigated the impact of several key components of the system on performance using extended core tasks in the NIST 2010 SRE that involved both microphone and telephone trials. Some important observations made from our experiments were as follows: 1) the NDA was found to be consistently more effective than the LDA for inter-session variability compensation in i-vector based speaker recognition, 2) the fMLLR based features provided better representation than raw MFCCs for matched data conditions (i.e., telephony trials), and 3) the DNN based UBM with large number of components (i.e., 10k senones) resulted in remarkable improvements in the performance of our system. To the best of our knowledge, the results presented in this paper represent the best performances reported to date on the extended core tasks in the NIST 2010 SRE.  



\balance
\bibliographystyle{IEEEtran}
\bibliography{refs_odyssey2016,IEEEabrv,IEEEfull}

%

\end{document}